\documentstyle[eqsecnum,floats,preprint,aps]{revtex}

\tighten

\def\al{\alpha}

\def\be{\begin{equation}}
\def\ee{\end{equation}}
\def\bea{\begin{eqnarray}}
\def\eea{\end{eqnarray}}

\newcommand{\vp}{\varphi}

\newcommand{\Lag}{{\cal L}}
\newcommand{\th}{\theta}

\newcommand{\ald}{{\dot{\alpha}}}

\newcommand{\sigb}[1]{\bar{\sigma}^{{#1}}}
\newcommand{\sgth}{\sigma^\vp}
\newcommand{\sgr}{\sigma^r}
\newcommand{\sgz}{\sigma^z}
\newcommand{\la}{\lambda}
\newcommand{\lab}{\bar{\lambda}}

\newcommand{\phb}{\bar{\phi}}
\newcommand{\xib}{\bar{\xi}}

\newcommand{\eth}[1]{e^{{#1}\vp}}
\newcommand{\spinU}{\mbox{\scriptsize 
		$\left(\begin{array}{c} 1 \\ 0 \end{array} \right)$}}

\begin{document}

\title{Supersymmetric Strings and Fermionic Zero-Modes\footnote{Talk
at COSMO-97, Ambleside, England, September 15-19 1997. To appear in the 
proceedings.}}

\author{Mark Trodden\footnote{trodden@theory1.phys.cwru.edu.}}

\address{~\\Particle Astrophysics Theory Group \\
Department of Physics \\
Case Western Reserve University \\
10900 Euclid Avenue \\
Cleveland, OH 44106-7079, USA.}
\maketitle

\begin{abstract}
I review recent work concerning the microphysics of cosmic string solutions 
of $N=1$ supersymmetric gauge field theories. 
\end{abstract}

\setcounter{page}{0}
\thispagestyle{empty}
\vfill
\baselineskip 14pt

\noindent CWRU-P13-97

\eject

\vfill

\eject

\baselineskip 24pt plus 2pt minus 2pt

\section{Introduction}
Topological soliton solutions of gauge field theories may have played an 
important role in the evolution of the universe. In addition to the 
well-known gravitational effects of such objects, the microphysics of 
topological defects can also have cosmological consequences. 

In light of continued theoretical and experimental interest in supersymmetry 
(SUSY), it seems natural to explore the properties of topological 
solitons in SUSY field theories.\cite{DDT1} In this short review I focus on 
such an analysis in a simple abelian model with one SUSY generator ($N=1$).

\section{A SUSY Abelian Higgs Model and Cosmic Strings}
This theory consists of
a vector superfield $V(A_{\mu}, \lambda, D)$ and three chiral superfields 
$\Phi_i(\phi_i, \psi_i, F_i)$, ($i=\pm,0$),
with $U(1)$ charges $q_i$. 
Here, $\phi_i$ are complex scalar fields and $A_\mu$ is a vector field. These
correspond to the familiar bosonic fields of the abelian Higgs model.
The fermions $\psi_{i \alpha}$, $\lab_{\alpha}$ and
$\la_{\alpha}$ are Weyl spinors. The gauge symmetry is broken through an 
$F-$term generated by the holomorphic superpotential 
$W(\Phi_i) = \mu \Phi_0 (\Phi_+ \Phi_- - \eta^2)$, 
with $\eta$ and $\mu$ real. In this model the potential is minimized for 
$F_i=0$ and $D=0$.

Setting all the fermion fields to zero, there exists a 
Nielsen-Olesen\cite{NO} cosmic string solution with ansatz

\begin{eqnarray}
\phi_+ & = & \phi_-^\ast = \eta e^{in\vp}f(r) \ , \\
A_\mu & = & -\frac{2}{g} n \frac{a(r)}{r}\delta_\mu^\vp \ , \\
F_0 & = & \mu \eta^2 (1 - f(r)^2) \ .
\label{StringSol}
\end{eqnarray}
All other
fields are zero and the profile functions obey the usual
differential equations of the abelian-Higgs model with boundary conditions 
$f(0)=a(0)=0$ and $\lim_{r\rightarrow 
\infty}f(r)=\lim_{r\rightarrow\infty}a(r)=1$.

\section{Fermion Zero Modes and SUSY Transformations}

The fermionic sector has Yukawa couplings

\begin{equation}
\Lag_Y = i\frac{g}{\sqrt{2}}
	 \left(\phb_+ \psi_+ - \phb_- \psi_-\right) \la
		 - \mu \left(\phi_0 \psi_+ \psi_- + \phi_+ \psi_0 \psi_- 
		+ \phi_- \psi_0 \psi_+ \right) + (\mbox{c.c.})
\end{equation}
and, as with a non-SUSY theory, non-trivial zero energy fermion
solutions can exist around the string. Consider the fermionic ansatz 
$\psi_i = \spinU \psi_i(r,\vp)$, $\la = \spinU \la(r,\vp)$.
If we can find solutions for the $\psi_i(r,\vp)$ and $\la(r,\vp)$ then, 
following Witten,\cite{Witten} we know that solutions of the form

\be
\Psi_i=\psi_i(r,\vp)e^{i\chi(z + t)} \ , \
 \Lambda=\la(r,\vp)e^{i\chi(z + t)} \ ,
\label{witteq}
\ee
with $\chi$ some function, represent left moving superconducting 
currents flowing along the string at the speed of light. Thus we must
solve for $\psi_i(r,\vp)$ and $\la(r,\vp)$. 

Consider performing an infinitesimal SUSY transformation with infinitesimal
Grassmann parameters $\xi$, $\xib$. This induces a gauge transformation and 
we undo this using a SUSY gauge parameter 
$\Lambda=ig\xib\sigb{\mu}\th A_\mu (y)$. Writing everything in terms of the 
background string fields and keeping only terms up to first order, the string
fields are unchanged and the fermion fields become

\begin{eqnarray}
\la_\al &\rightarrow& \frac{2na'}{gr}i(\sgz)^\beta_\al \xi_\beta \ , \\
(\psi_\pm)_\al &\rightarrow& \sqrt{2} \left(if'\sgr \mp
\frac{n}{r}(1-a)f \sgth\right)_{\al \ald} \xib^\ald \eta \eth{\pm in} \ , \\
(\psi_0)_\al &\rightarrow& \sqrt{2}\mu\eta^2(1-f^2)\xi_\al \ ,
\end{eqnarray}
where $\sgr$ and $\sgth$ are the cylindrical Pauli matrices.

If we choose $\xi_2 = 0$ and $\xi_1 = -i\delta/(\sqrt{2}\eta)$, where $\delta$
is a complex constant, we obtain

\begin{eqnarray}
\la_1 & = & \delta\frac{n\sqrt{2}}{g\eta}\frac{a'}{r} \ , \\ 
(\psi_\pm)_1 & = & \delta^\ast 
\left[f'\pm \frac{n}{r}(1-a)f\right]\eth{i(\pm n-1)} \ , \\
(\psi_0)_1 & = & -i\delta\mu\eta(1-f^2) \ .
\end{eqnarray}
It is these fermion solutions which are responsible for the string
superconductivity.
Similar expressions can be found when $\xi_1 = 0$. Note that these solutions 
can be checked in the special case of the Bogomolnyi limit and they agree 
with those found by previous authors.\cite{Garriga}

\section{Concluding Remarks}
Supersymmetric Abelian-Higgs models naturally possess cosmic string solutions 
which carry fermionic supercurrents. This result also holds in nonabelian 
models where the structure is considerably richer.\cite{toappear}

In a cosmological context such strings may settle into vorton states and
thus may be cosmologically catastrophic. A potential solution is that
it appears that the zero modes responsible for superconductivity vanish when
supersymmetry is broken.\cite{toappear} 

Finally, simple potentials of the type described here can also give rise
to a separate cosmological regime of hybrid inflation.
When such inflation ends the strings form. In such a scenario density 
fluctuations from both the inflationary epoch and the defect dynamics should
be taken into account.

\section*{Acknowledgments}

I would like to thank the conference organisers for putting together such an
enjoyable meeting and in particular for the focus on {\it particle} cosmology.
I also owe thanks to my collaborators, Anne-Christine Davis and 
Stephen Davis. This work was supported by the Department of Energy 
(D.O.E.), the National Science Foundation (N.S.F.) and by funds provided
by Case Western Reserve University.

\end{document}